\documentclass{article}

\usepackage[english]{babel}

\usepackage[letterpaper,top=2cm,bottom=2cm,left=3cm,right=3cm,marginparwidth=1.75cm]{geometry}

\usepackage{amsmath}
\usepackage{amssymb}
\usepackage{graphicx}
\usepackage{xcolor}
\usepackage[colorlinks=true, allcolors=blue]{hyperref}
\usepackage{amsthm}
\usepackage{authblk}
\usepackage[compress,square,comma,numbers]{natbib}

\title{Quasinormal modes of Reissner-Nordstr\"om--AdS: the approach to extremality}
\date{}

\author[1,2]{Filip Ficek\footnote{\href{mailto:filip.ficek@univie.ac.at}{filip.ficek@univie.ac.at}}}
\author[3,4]{Claude Warnick}
\affil[1]{Faculty of Mathematics, University of Vienna, Oskar-Morgenstern-Platz 1, 1090 Vienna, Austria}
\affil[2]{Gravitational Physics Group, University of Vienna, W\"{a}hringer Stra{\ss}e 17, 1090 Vienna, Austria}
\affil[3]{Department of Pure Mathematics and Mathematical Statistics, University of Cambridge, Wilberforce Road, Cambridge, CB3 0WB, United Kingdom}
\affil[4]{Department of Applied Mathematics and Theoretical Physics, University of Cambridge, Wilberforce Road, Cambridge, CB3 0WB, United Kingdom}

\begin{document}
\maketitle

\begin{abstract}
We consider the quasinormal spectrum of scalar and axial perturbations of the Reissner-Nordstr\"om--AdS black hole as the horizon approaches extremality. By considering a foliation of the black hole by spacelike surfaces which intersect the future horizon we implement numerical methods which are well behaved up to and including the extremal limit and which admit initial data which is nontrivial at the horizon. As extremality is approached we observe a transition whereby the least damped mode ceases to be oscillatory in time, and the late time signal changes qualitatively as a consequence.
\end{abstract}

\section{Introduction}

Numerical \cite{Buo, Berti} and observational \cite{LIGO} evidence shows that a black hole spacetime will, in response to a perturbation, produce radiation at (complex) frequencies which are characteristic of the black hole. These frequencies are the quasinormal frequencies, and to each such frequency is associated a quasinormal mode -- a solution of a linear equation on the black hole background, satisfying suitable boundary conditions at any horizons and (if relevant) at null infinity \cite{Kokkotas:1999bd, Berti:2009kk, Konoplya:2011qq}.

In recent years, a satisfactory mathematical understanding of the quasinormal modes of subextremal de Sitter black hole spacetimes has developed, starting with results for Schwarzschild-de Sitter \cite{SZ, BH},  culminating in a general theory for de Sitter black holes \cite{Vasy} including the full subextremal range of Kerr--de Sitter black holes \cite{Dyat, PV}. In the subextremal anti-de Sitter setting, analogous results have been shown \cite{Me, Gannot:2016awr}. For a thorough overview see \cite{DZ}, and for an explicit worked example using this approach, see \cite{Bizon:2020qnd}.

The majority of the works cited in the previous paragraph impose regularity at the future horizon(s) in order to characterise the quasinormal modes, an approach which in the physics literature goes back to Schmidt \cite{Schmidt}. It can be shown that time-harmonic solutions to the linearised equations which extend smoothly\footnote{In fact analytically \cite{GZ2, PV2}} across the future horizons exist only for a discrete set of complex frequencies, which can be identified with the quasinormal frequencies. Regularity at the horizon plays the role of `in/outgoing' boundary conditions in more traditional treatments. This approach breaks down when the black hole horizon is extremal or the spacetime is asymptotically flat. (For the purposes of our discussion, an asymptotically flat end may be thought of as the extremal limit of a subextremal cosmological horizon, and when we refer to a subextremal spacetime we implicitly assume that it has no asymptotically flat ends). 

The behaviour of the quasinormal spectrum as a spacetime approaches extremality has been the topic of significant interest in the physics literature \cite{Hod:2008, Hod:2008zz, Hod:2009td, Hod:2010hw, Hod:2011zzd, Hod:2012, Yang:2013uba, Zimmerman:2015trm}. In particular, going back at least to Detweiler \cite{Detweiler:1980gk}, two distinct behaviours have been observed for the quasinormal frequencies as the surface gravity, $\kappa$ approaches zero. Firstly, it appears that a generic feature of near-extremal spacetimes is the existence of a sequence of `zero damped modes' with damping rates approximately $n\kappa$, $n=1, 2, 3, \ldots$, which accumulate at some given frequency in the limit $\kappa\to0$. On the other hand, in certain regions of the complex frequency plane, the quasinormal frequencies are largely unaffected by the extremal limit -- the frequencies settle down to limiting values without accumulating (called `damped modes'). 

In \cite{Gajic:2019qdd} the second author, together with Gajic, considered the Reissner-Nordstr\"om-de Sitter black hole and showed that in the limit where both horizons become extremal there is a sector in the complex plane in which, away from the origin, only the damped mode behaviour is observed. In \cite{Joykutty:2021fgj} Joykutty established the existence of purely damped modes in several situations involving a horizon approaching extremality, including that of the Reissner--Nordstr\"om-de Sitter black hole with either horizon becoming extremal (see also \cite{Hintz:2021vfl, Hintz:2021rbv} for a closely related result).

In this paper, we aim to study numerically the quasinormal spectrum of the Reissner-Nordstr\"om--anti-de Sitter black hole in the approach to extremality. Our approach is motivated by that taken in \cite{Gajic:2019oem, Gajic:2019qdd}, and involves working in coordinates which are regular at the horizon (see \cite{GalZw} for an alternative approach). We choose a foliation by spacelike surfaces as this most easily can be adapted to the case of multiple horizons, but a null (or mixed null/spacelike) slicing could also be considered and should give similar results. The quasinormal spectrum of the Reissner-Nordstr\"om--anti-de Sitter black hole has been studied in \cite{Ber03}, but the parameter ranges considered in that paper do not include a neighbourhood of the extremal case. We extend their results for scalar and axial perturbations all the way to extremality.

We work both in the time and frequency domains, enabling cross-checking of results between independent computations. In the time domain our choice of slicing permits us to simulate time evolution up to and including the horizon, without introducing any artificial boundaries. In the frequency domain we use a modified Leaver method to determine quasinormal frequencies (this was demonstrated to correctly locate the quasinormal frequencies in \cite{Gajic:2019oem}). In both cases our methods are designed such that there is no degeneration as $\kappa \to 0$ and so that we are able to consider initial data which is non-trivial at the future event horizon.

A particularly interesting feature we observe for both the conformal wave equation and the perturbations is a threshold in the black hole parameter space at which the qualitative behaviour of the fields changes. This happens when the surface gravity is sufficiently small that the slowest decaying purely damped mode becomes the dominant late-time behaviour. On one side of this threshold the late time behaviour is oscillatory, but closer to extremality the dominant behaviour becomes pure exponential decay. In the extremal limit, this ever-slower exponential decay becomes the polynomial decay expected for an extremal black hole \cite{Zimmerman:2016qtn}.

After this brief introduction, in Section \ref{sec:toymodel} we briefly consider a scalar toy equation which can be solved explicitly in terms of special functions in order to verify our numerical methods, before moving on to study the conformal wave equation and the scalar and axial perturbations for the Reissner-Nordstr\"om--anti-de Sitter black hole as the horizon approaches extremality in Section \ref{sec:rnads}. Our results are summarised in Section \ref{sec:conclusions}

\section{Explicitly solvable toy-model}\label{sec:toymodel}

\begin{figure}
    \centering
    \includegraphics[scale=0.5]{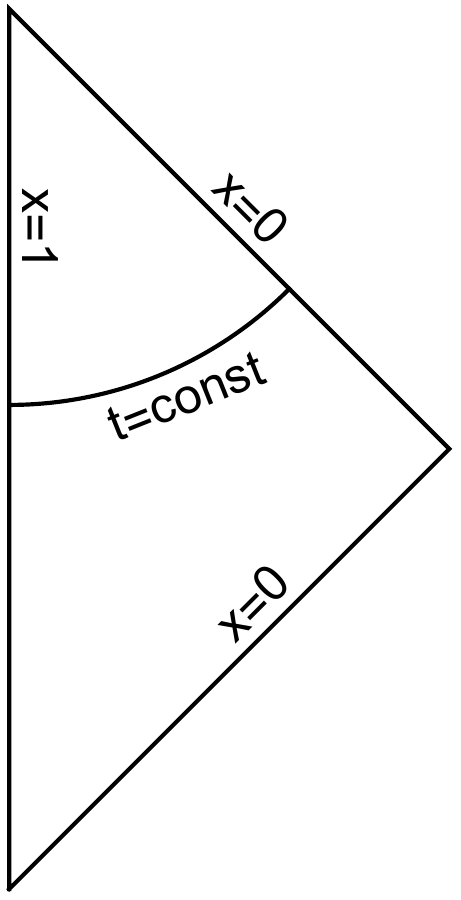}
    \caption{Causal diagram for the spacetime described by Eq.\ (\ref{eqn:toymodel_g}). The angular dimension is suppressed.}
    \label{fig:toymodel}
\end{figure}

We consider the following equation which models the behaviour of a wave exterior to an extremal AdS black hole (throughout the whole article we use the geometrised unit system $c=G=1$):
\begin{equation}\label{eqn:toymodel}
    -\partial_t^2\psi+\partial_x(x^2\partial_x\psi)+2\partial_t\partial_x\psi+\partial_\theta^2\psi+\frac{1}{4}\psi=0.
\end{equation}
where $(t,x,\theta)\in[0,T)\times [0,1]\times S^1$. The principle part of this operator agrees with that of the wave operator for the spacetime with metric $g$ given by
\begin{equation}\label{eqn:toymodel_g}
    g^{-1}=-\frac{\partial}{\partial t} \otimes \frac{\partial}{\partial t} +  \frac{\partial}{\partial t}\otimes \frac{\partial}{\partial x}+\frac{\partial}{\partial x}\otimes \frac{\partial}{\partial t}+ x^2 \frac{\partial}{\partial x}\otimes \frac{\partial}{\partial x} + \frac{\partial}{\partial \theta}\otimes \frac{\partial}{\partial \theta}
\end{equation}
The causal diagram of this spacetime is shown in Fig.\ \ref{fig:toymodel}. It enjoys the presence of an extremal horizon at $x=0$, as can be seen from the fact that $g(\partial_t, \partial_t)$ behaves like $x^2$ in its vicinity. Let us point out that the time coordinate variable is chosen in such a way that the surfaces of constant time $t$ penetrate this horizon and the metric extends smoothly across $x=0$.

In this section we investigate solutions to Eq.\ (\ref{eqn:toymodel}) satisfying the Dirichlet condition $\psi(t,1,\theta)=0$ at $x=1$. At $x=0$ we do not require a boundary condition, owing to the presence of the horizon. We assume that initial conditions $\psi(0, x, \theta)$ and $\partial_t \psi(0, x, \theta)$ are specified. 

According to the prescription of \cite{Gajic:2019oem, Gajic:2019qdd}, in order to find the quasinormal frequencies of Eq.\ (\ref{eqn:toymodel}) we should seek solutions to the equation of the form
\[
\psi(t, x, \theta) = e^{s t + i m \theta} u(x)
\]
which satisfy the boundary condition at $x=1$ and which have improved regularity (relative to a generic solution) at $x=0$. Here we have made use of the rotational symmetry to restrict attention to a single angular mode. The function $u$ can be seen to satisfy the following ODE
\begin{align}\label{eqn:ode}
    -\frac{d}{dx}\left(x^2\frac{d}{dx}u\right)-2s\frac{d}{dx}u+\left(s^2+m^2-\frac{1}{4}\right)u=0.
\end{align}
Introducing a function $v$ given by
\begin{align*}
    u(x)=e^{\frac{s}{x}}\sqrt{\frac{s}{x}} v\left(\frac{s}{x}\right),
\end{align*}
the left hand side of Eq.\ (\ref{eqn:ode}) becomes the modified Bessel equation
\begin{align*}
    z^2 \frac{d^2 v}{dz^2}+z \frac{dv}{dz}-\left(z^2+\lambda^2\right)v=0,
\end{align*}
where $z=\frac{s}{x}$ and $\lambda=\sqrt{s^2+m^2}$. Thus, the general solution of (\ref{eqn:ode}) can be written as
\begin{align}\label{eqn:solu}
    u(x)=e^{\frac{s}{x}}\sqrt{\frac{s}{x}}\left(a\, K_\lambda\left(\frac{s}{x}\right)+b\, I_\lambda\left(\frac{s}{x}\right)\right),
\end{align} 
where $a, b \in \mathbb{C}$ and $I_{\lambda}(z), K_\lambda(z)$ are the modified Bessel functions of the first and second kind respectively \cite{Watson}.

In order to discuss the regularity condition that we impose at $x=0$, recall that a smooth function $f: (0, 1) \to \mathbb{C}$ is  $(\sigma, k)-$Gevrey regular with $\sigma, k>0$ if there exists a constant $C$ such that
\[
\sup_{x \in (0,1)} \left|f^{(n)} (x) \right| \leqslant C \sigma^{-n} (n!)^k.
\]
Note that for fixed $k$, increasing $\sigma$ imposes a more stringent regularity condition on $f$.

By a careful analysis of the asymptotic series of the modified Bessel functions using the approach of \cite[\S7.31]{Watson}, together with \cite[Prop 8]{Bas}  and \cite[eqn (A.1)]{GalZw} it is possible to show that for fixed $s$ with $|\arg s|<\pi$:
\begin{itemize}
    \item $e^{\frac{s}{x}}\sqrt{\frac{s}{x}}\, K_\lambda\left(\frac{s}{x}\right)$ is $(\sigma, 2)-$Gevrey regular for any $\sigma <|s|$
    \item $e^{\frac{s}{x}}\sqrt{\frac{s}{x}}\, I_\lambda\left(\frac{s}{x}\right)$ is \textit{not} $(\sigma, 2)-$Gevrey regular if $\left \{ \begin{array}{ll}
        \sigma>0,  & |\arg s|\leqslant \pi/2 \\
         \sigma> |s|-|\Im(s)|,\,\, &  \pi/2<|\arg s|<\pi
    \end{array} \right.$
\end{itemize}
In particular, this implies that we should make the choice $b=0$ in \eqref{eqn:solu} to single out the more regular branch of solutions (in the Gevrey sense) at $x=0$. In order that our solution also satisfies the boundary condition at $x=1$ we require
\begin{align}\label{eqn:spole}
    K_{\sqrt{s^2+m^2}}(s)=0.
\end{align}
Thus the quasinormal frequencies are precisely the solutions to Eq.\ \eqref{eqn:spole}. Since $K_{\lambda}(z)$ is an entire, even, function of $\lambda$, the branch points at $\pm im$ are removable, however, a branch point at $s=0$ will be present in general.

\begin{figure}
    \centering
    \includegraphics[scale=0.5]{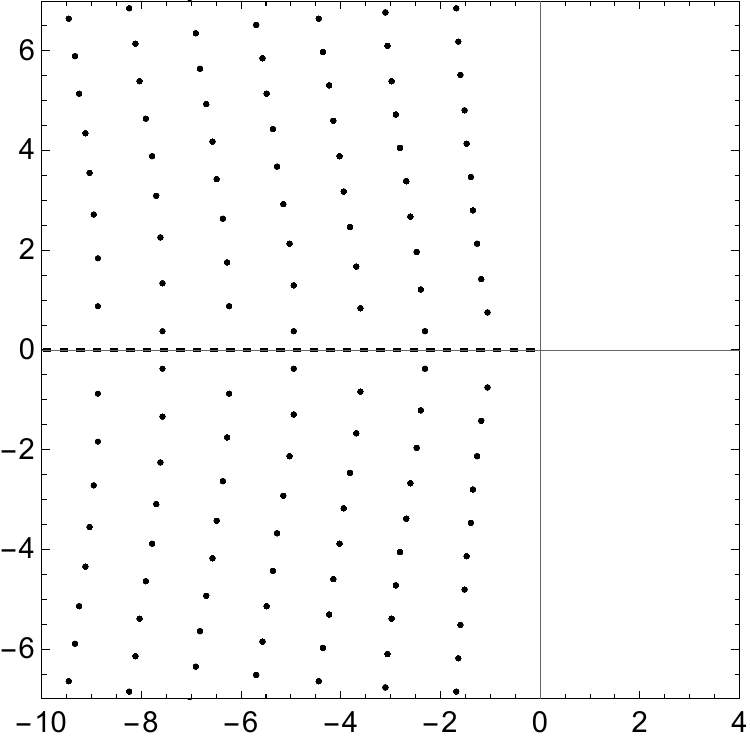}
    \caption{Locations of the lowest solutions to Eq.\ (\ref{eqn:spole}) in the complex plane. The dashed line represents the branch cut.}
    \label{fig:poles}
\end{figure}

The solutions of Eq.\ \eqref{eqn:spole} for various $m$ are given in Table \ref{tab:qnfOG} and presented in Fig.\ \ref{fig:poles}. As well as the locations of the quasinormal frequencies, we also obtain an explicit formula for the corresponding quasinormal modes:
\begin{align*}
    \psi(t, x, \theta) =a\, e^{st+i m \theta +\frac{s}{x}}\sqrt{\frac{s}{x}} K_{\sqrt{s^2+m^2}}\left(\frac{s}{x}\right),
\end{align*}
where $s$ is a solution to Eq.\ (\ref{eqn:spole}) and $a$ is any constant.

\begin{table}
\centering
\begin{tabular}{r|rrrrr}
$m=0$ &  &&&& \\
$m=1$ &  &&&& \\
$m=2$ & $-1.060\pm 0.736i$ &&&& \\
$m=3$ & $-1.167\pm 1.441i$ & $-2.302\pm 0.371i$ &&& \\
$m=4$ & $-1.258\pm 2.122i$ & $-2.382\pm 1.216i$ &&& \\
$m=5$ & $-1.337\pm 2.798i$ & $-2.494\pm 1.965i$ & $-3.624\pm 0.842i$&& \\
$m=6$ & $-1.407\pm 3.473i$ & $-2.603\pm 2.675i$ & $-3.709\pm 1.685i$&$-4.934\pm 0.382i$& \\
$m=7$ & $-1.471\pm 4.149i$ & $-2.705\pm 3.367i$ & $-3.819\pm 2.457i$&$-4.951\pm 1.310i$& \\
$m=8$ & $-1.529\pm 4.826i$ & $-2.800\pm 4.053i$ & $-3.931\pm 3.191i$&$-5.038\pm 2.151i$&$-6.239\pm 0.869i$ \\
\end{tabular}
\caption{\label{tab:qnfOG}Approximate quasinormal frequencies for Eq.\ (\ref{eqn:toymodel}).}
\end{table}

\begin{figure}
\includegraphics[width=0.45\textwidth]{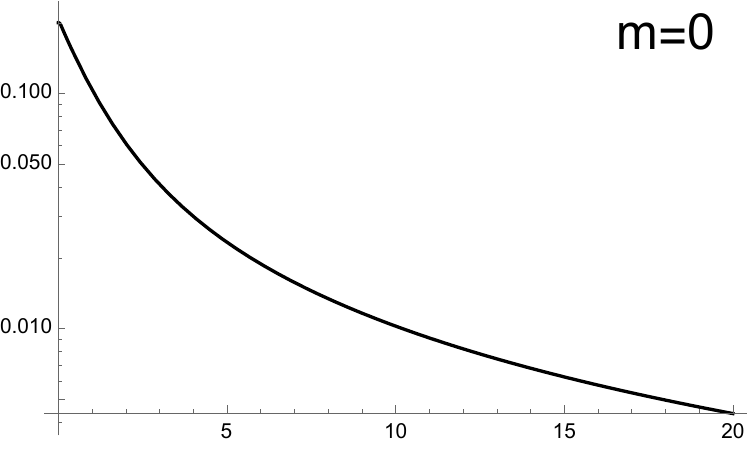}
~
\includegraphics[width=0.45\textwidth]{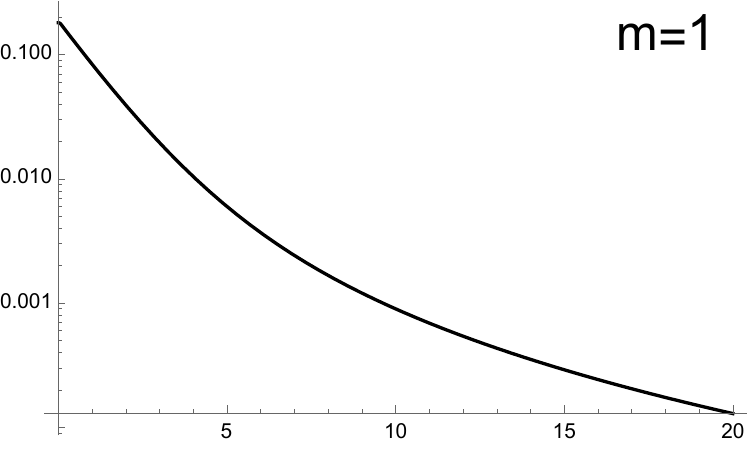}
\\
\includegraphics[width=0.45\textwidth]{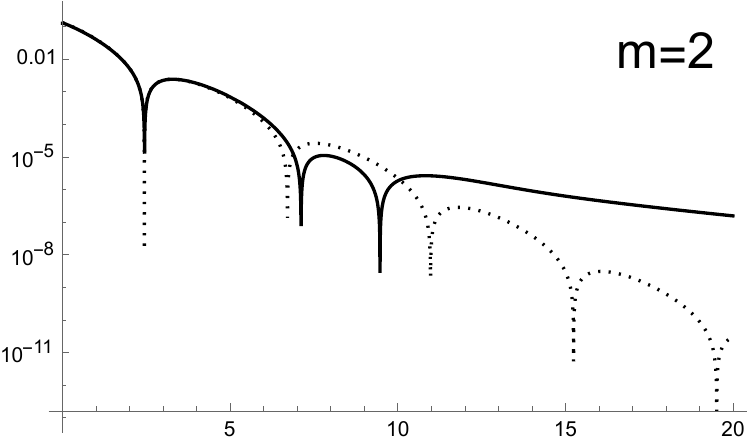}
~
\includegraphics[width=0.45\textwidth]{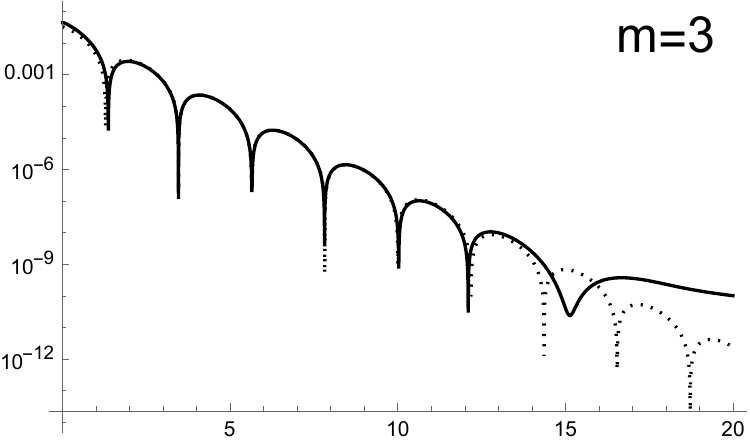}
\\
\includegraphics[width=0.45\textwidth]{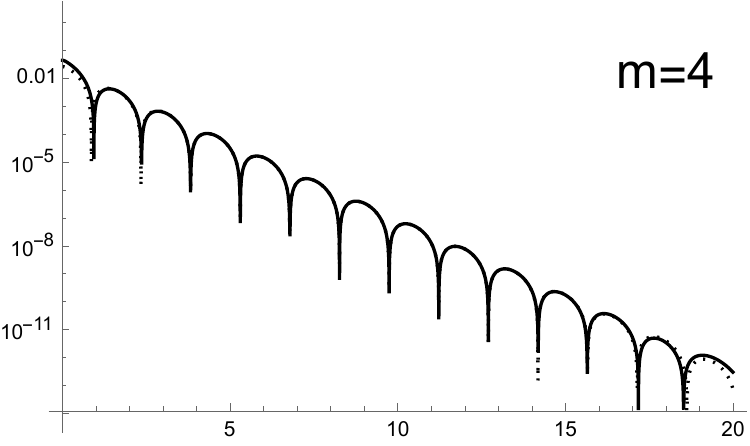}
~
\includegraphics[width=0.45\textwidth]{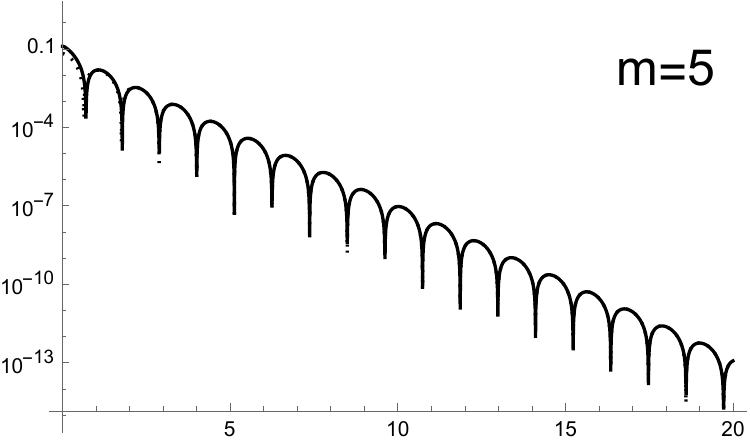}
\caption{Plots of solutions to Eq.\ (\ref{eqn:toymodel}) for the initial data $\psi(0,x,\theta)=(1-x)\cos m\theta$, $\psi_t(0,x,\theta)=0$ with various $m$. The solid lines show functions $|\psi(t,0.8,\frac{\pi}{7})|$ and the dotted lines represent the fitted dominating quasinormal mode.}
\label{fig:genLin}
\end{figure}

The presented results can be confronted with the numerical approach. Equation (\ref{eqn:toymodel}) can be solved with the use of the pseudospectral scheme \cite{NumRec}. Since at $x=1$ we impose the Dirichlet condition, to control it we employ the Gauss-Radau quadratures \cite{NumRec}. The solutions that we are looking for are decaying exponentially with time in the quasinormal regime. Hence, to improve the precision of the scheme one can evolve an auxiliary function $\tilde{\psi}(t,x,\theta)=e^{\alpha t} \psi(t,x,\theta)$ with $\alpha>0$ being a suitably chosen constant. The accuracy of this method can be controlled by the energy
\begin{equation*}
    E(t)=\frac{1}{2}\int_0^1\int_0^{2\pi}\left[|\partial_t\psi|^2+x^2|\partial_x\psi|^2+|\partial_\theta\psi|^2-\frac{1}{4} |\psi|^2\right]dx\,d\theta.
\end{equation*}
One can use the Hardy inequality to show that it is positive. Due to the absence of term $\partial_t \psi$ in Eq.\ (\ref{eqn:toymodel}) and the coefficient next to $\partial_{\tau x}$ being constant, this energy changes in time only via the leakage through the horizon. This change is given by a simple expression involving only an integration over the angular variable
\begin{equation*}\label{eqn:deltaE}
    E'(t)=-\int_0^{2\pi}|\partial_t\psi|^2_{x=0}\,d\theta.
\end{equation*}
As can be seen in Fig.\ \ref{fig:genLin} for larger values of $m$ one can easily observe the quasinormal regime. The quasinormal frequencies obtained via fitting agree with the lowest values from Table \ref{tab:qnfOG}. Note that the initial data is chosen to be non-zero at the horizon -- a key feature of the approach of \cite{Gajic:2019oem, Gajic:2019qdd} is that such initial data is permissible.

\begin{figure}
\centering
\includegraphics[width=0.45\textwidth]{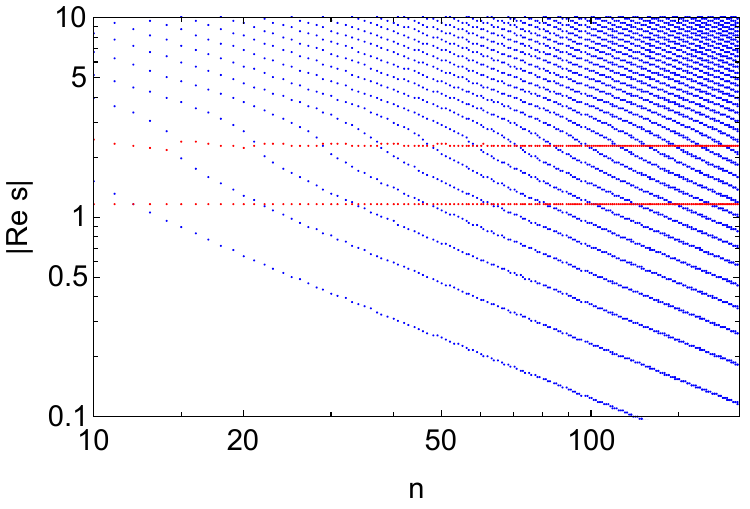}
\caption{Log-log plot of absolute values of real parts of solutions to equation $H_n=0$ for various values of $n$, where $H_n$ is given by the recurrence relation (\ref{eqn:Hk_rel}). Blue dots show spurious solutions that are purely real, while the red ones have non-trivial imaginary part and converge to the quasinormal frequencies.}
\label{fig:conv_toy}
\end{figure}

An alternative approach to finding the quasinormal modes frequencies for our toy-model is by the Leaver method \cite{Lea85,Lea90}. Let us fix some angular number $m$ and again look for solutions of the form $\psi(t,x,\theta)=e^{st}e^{im \theta} u(x)$. Then $u$ satisfies Eq.\ (\ref{eqn:ode})
and we can expand it into a Taylor series around $x=1$: $u(x)=\sum_{k=0}^{\infty}H_k (1-x)^k$. The coefficients $H_k$ must satisfy the following recurrence relation
\begin{equation}\label{eqn:Hk_rel}
    H_{k}=\frac{1}{(k-1)k}\left[2(k-1)(k-1+s)H_{k-1}-\left((k-2)(k-1)-\left(s^2+m^2-\frac{1}{4}\right)\right)H_{k-2}\right]
\end{equation}
for $k\geq 2$. Since we impose the Dirichlet condition at $x=1$, we need to set $H_0=0$. The regularity condition at $x=0$ suggests that $H_k$ should converge to zero as $k\to\infty$. It gives us a quantization condition on $s$. One can obtain the appropriate values of $s$ using the method of continued fractions, but in our case it is enough to assume that for some sufficiently large value of $n$ one has $H_n=0$ (in none of the cases considered in this article the continued fraction method led to significantly faster convergence). It leads to a polynomial, whose zeroes include approximations to $s$ we seek and many superfluous values. To identify the correct values of $s$ one can change $n$ and see which roots converge. It is presented in Fig.\ \ref{fig:conv_toy}. From the analytical solution to the toy model we know that proper quasinormal frequencies have non-zero imaginary part (red dots in the plot). Quasinormal frequencies obtained with this method agree with the ones resulting from Eq.\ (\ref{eqn:spole}).

\section{Reissner-Nordstr\"{o}m-anti-de Sitter black hole}\label{sec:rnads}
Now we would like to apply the same methods to study the quasinormal modes in the Reissner-Nordstr\"{o}m-anti-de Sitter (RNAdS) spacetime. Let us start by investigating the wave operator in this spacetime. Our first step consists of finding a suitable coordinate system in which it becomes similar to the one from the toy model.

\begin{figure}
    \centering
        \includegraphics[width=0.45\textwidth]{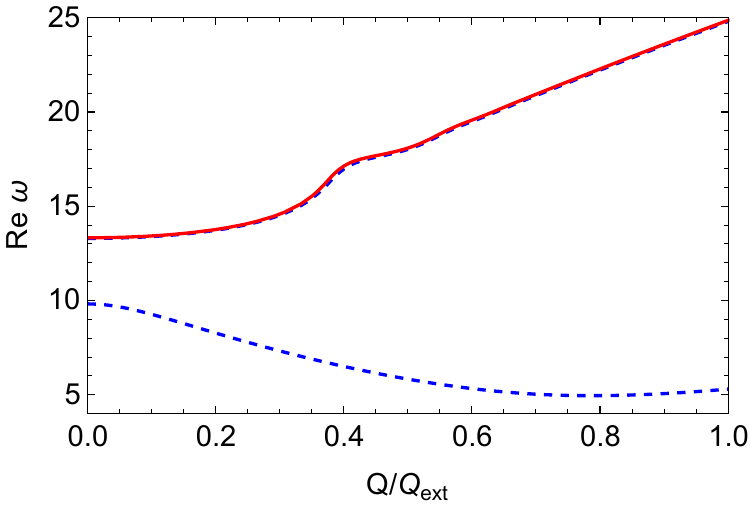}\qquad
    \includegraphics[width=0.45\textwidth]{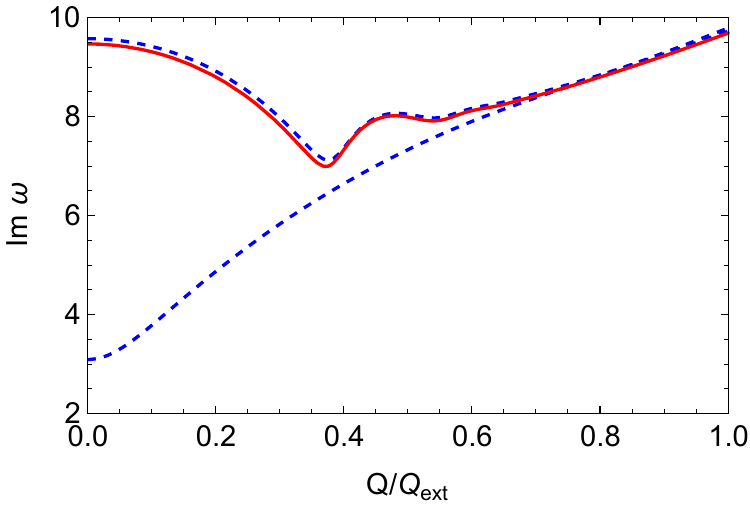}
    \caption{Quasinormal frequencies for perturbations of Reissner-Nordstr\"{o}m-anti-de Sitter spacetime with $\ell=1$ and $r_H=5$. The solid red line denotes scalar perturbations with $l=0$ while blue dashed lines shows results for axial perturbations with $l=2$.}
    \label{fig:berti}
\end{figure}

In spherical cooridnates $(t,r,\theta,\phi)$ the line element of the RNAdS spacetime is given by
\begin{equation}\label{eqn:RNAdS0}
    ds^2=-\left(1-\frac{2M}{r}+\frac{Q^2}{r^2}+\frac{r^2}{\ell^2}\right)dt^2+\left(1-\frac{2M}{r}+\frac{Q^2}{r^2}+\frac{r^2}{\ell^2}\right)^{-1} dr^2+r^2 d\Omega^2,
\end{equation}
where $d\Omega^2$ is a line element on a two-dimensional unit sphere. The values of $M$ and $Q$ are interpreted as the mass and the charge of the black hole, respectively, while $\ell$ gives a specific length-scale connected with the cosmological constant $\Lambda$ via $\Lambda=-3/ \ell^2$. In the generic case this spacetime has two spherical horizons called the Cauchy horizon (of a radius $r_C$) and the event horizon (of a radius $r_H$). However, if $Q=0$, the latter vanishes and we get a Schwarzschild-anti-de Sitter spacetime. On the other hand, for $Q$ large enough these horizons coincide, i.e., $r_H=r_C=(3M-\sqrt{9M^2-8Q^2})/2$ and then their position does not depend on the cosmological constant. This situation is called an extremal case, in contrast to the regular case in which $r_C<r_H$. In the following we want to cover both regular and extremal cases so we need a framework that will suitably handle both possibilites. For this purpose it is convenient to introduce the following quantities. Let $\rho=r/r_H$ be a new radial variable and $t_H=t/r_H$ a new temporal variable. We also define the parameters $\sigma=r_C/r_H$, and $\lambda=r_H^2/\ell^2$. Then Eq.\ (\ref{eqn:RNAdS0}) can be written as
\begin{equation}\label{eqn:RNAdS}
    ds^2=r_H^2 \left[ -f(\rho) dt_H^2+f(\rho)^{-1} d\rho^2+\rho^2 d\Omega^2 \right]
\end{equation}
where
\begin{equation*}
    f(\rho)=1-\frac{1+\sigma}{\rho}\left(1+\left(1+\sigma^2\right)\lambda\right)+\frac{\sigma}{\rho^2}\left(1+\left(1+\sigma+\sigma^2\right)\lambda\right)+\lambda \rho^2.
\end{equation*}
In this parametrisation $\sigma=1$ gives the extremal case, $\sigma=0$ represents the black hole with no charge, and $\lambda=0$ is the case with no cosmological constant. One can easily switch between parameters $(M,Q,\ell)$ and $(r_H,\sigma,\lambda)$ using the following relations
\begin{align*}
    \begin{cases}
    2M&=(1+\sigma)\left(1+\lambda(1+\sigma^2)\right)r_H,\\
    Q^2&=\sigma\left(1+\lambda(1+\sigma+\sigma^2)\right)r_H^2,\\
    \ell^2& =\lambda^{-1}r_H^2.
    \end{cases}
\end{align*}
Let us point out that $r^2_H$ takes a role of a scale factor in Eq.\ (\ref{eqn:RNAdS}) so from now on we assume $r_H=1$. For other values of $r_H$ one needs to perform elementary rescalings to recover appropriate results (as we do when plotting Fig.~\ref{fig:berti} in order to compare our results with \cite{Ber03}).

Next, we introduce a new time coordinate $\tau$ defined by $dt_H=d\tau+h'(\rho)\, d\rho$. The function $h$ here is chosen in such a way that the surfaces of constant $\tau$ cross the horizon, the new coordinate $\tau$ behaves like $t_H$ as $\rho\to\infty$, and the resulting wave operator behaves sufficiently well near the horizon. The last condition in fact means that the combination $f(\rho)h'(\rho)^2-f(\rho)^{-1}$, being the coefficient next to $\partial_{\tau}^2$ in the wave operator, does not blow up as $\rho\to 1$. This point is a little bit more subtle since we want to cover both regular (where $f$ behaves like $(\rho-1)$ near $\rho=1$) and extremal (where this behaviour is quadratic) cases with a single framework. It turns out that these conditions are satisfied by a function
\begin{equation*}
    h(\rho)=\frac{1}{(1-\sigma)\left(1+\lambda(3+2\sigma+\sigma^2)\right)}\log \left(\frac{\rho-1}{\rho}\right)-\frac{\sigma^2}{(1-\sigma)\left(1+\lambda(1+2\sigma+3\sigma^2)\right)}\log \left(\frac{\rho-\sigma}{\rho}\right).
\end{equation*}
Finally, we compactify the spatial domain by introducing new coordinate $x$ given by
\begin{equation}\label{eqn:xa}
    x=\frac{\rho-1}{\rho+a},
\end{equation}
where $a$ is some fixed nonnegative number and its choice will be discussed later. As a result, in these new coordinates the spacetime has a horizon at $x=0$ and an infinity is compactified to $x=1$, similarly to our toy-model. The metric then is given by
\begin{align}\label{eqn:metric_g}
    g=-f(\rho)\, d \tau^2 -2f(\rho) h'(\rho) \frac{1+a}{(1-x)^2} \, d\tau \, dx+\left(\frac{1}{f(\rho)}-f(\rho)h'(\rho)^2\right)\frac{(1+a)^2}{(1-x)^4}\, dx^2+\frac{(1+a x)^2}{(1-x)^2}\, d\Omega^2,
\end{align}
where $\rho$ inside the functions $f$ and $h$ needs to be replaced by $\rho=(1+a x)/(1-x)$.

For the sake of simplicity let us focus for a moment on the conformally invariant equation $\Box_g \psi -\frac{1}{6}R_g \psi=0$ \cite{Wald}. The wave operator $\Box_g$ resulting from our metric (\ref{eqn:metric_g}) contains a non-zero $\partial_\tau$ derivative term. It can be removed by employing the conformal invariance: one can check that the conformal transformation $(g,\psi)\to (\Omega^2 g,\Omega^{-1}\psi)$ with
\begin{equation*}
    \Omega(x)^2=\frac{(1-x)^2}{(1+a x)^2\, f\left(\frac{1+a x}{1-x}\right) h'\left(\frac{1+a x}{1-x}\right)}
\end{equation*}
leads to $\Box_{\Omega^2g}$ with no $\partial_\tau$ terms. The final step that needs to be done to get a problem similar to Eq.\ (\ref{eqn:toymodel}) is to fold the spatial derivatives $\partial_x$ and $\partial_x^2$ into a single expression. It can be achieved by simply dividing the whole equation $\Box_{\Omega^2 g}  \psi -\frac{1}{6}R_{\Omega^2 g} \psi=0$ by an appropriate integrating factor:
\begin{align*}
p(x)=\frac{(1+ax)^2 f\left(\frac{1+a x}{1-x}\right)^2 h'\left(\frac{1+a x}{1-x}\right)^2}{1+a}.
\end{align*}
Then, the resulting equation can be written as
\begin{equation*}\label{eqn:conformal_RNAdS_omega}
    a_{\tau\tau}(x) \partial_{\tau}^2 \psi+a_{\tau x}(x)\partial_\tau \partial_x \psi +\partial_x \left( a_{xx}(x)\partial_x \psi \right)+a_\omega(x)\left(\frac{1}{\sin\theta} \partial_\theta (\sin\theta \partial_\theta\psi)+ \frac{1}{\sin^2\theta}\partial_\phi^2 \psi\right)+a_0(x) \psi=0.
\end{equation*}
The dependence on the angular dimensions can be factored out with the help of the spherical harmonics $Y_{l,m}$ eventually leading to
\begin{equation}\label{eqn:conformal_RNAdS}
    a_{\tau\tau}(x) \partial_{\tau}^2 \psi+a_{\tau x}(x)\partial_\tau \partial_x \psi +\partial_x \left( a_{xx}(x)\partial_x \psi \right)+\left[a_0(x)-l(l+1)a_\omega(x) \right] \psi=0.
\end{equation}
The coefficients for general parameters $\sigma$ and $\lambda$ are rather complicated so we do not provide them explicitly. Instead, we note that for every $\lambda>0$ and $0\leq \sigma \leq 1$ we have $a_{\tau\tau}<0$ and $a_{\tau x}$ is a negative constant. In regular cases ($\sigma<1$) the coefficient $a_{xx}(x)$ behaves like a linear function near $x=0$, while for extremal charge ($\sigma=1$) this behaviour is quadratic, similarly to the toy-model (\ref{eqn:toymodel}).

\begin{figure}
    \centering
    \includegraphics[width=0.45\textwidth]{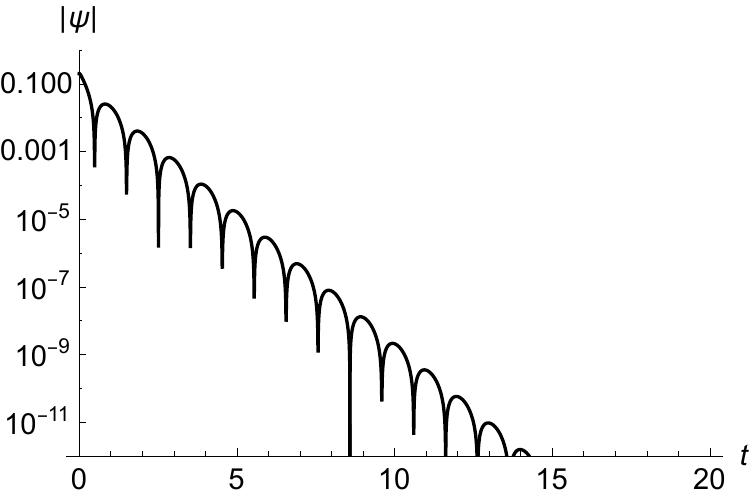}
    \includegraphics[width=0.45\textwidth]{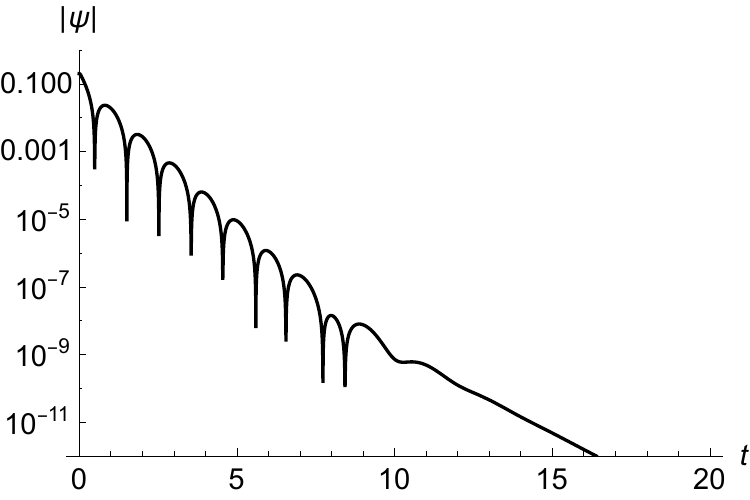}\\
    \includegraphics[width=0.45\textwidth]{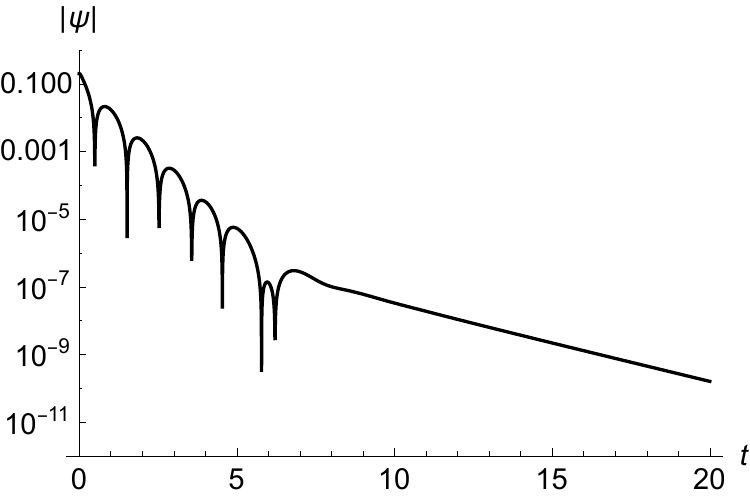}
    \includegraphics[width=0.45\textwidth]{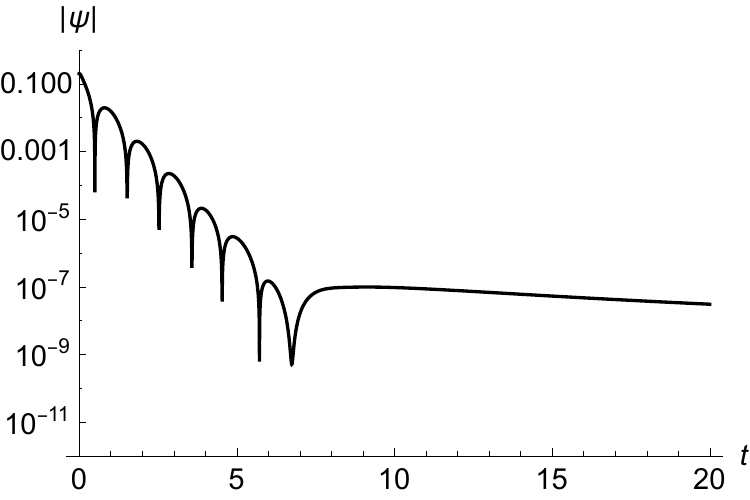}
    \caption{Plots of $|\psi|$ at $x=0.8$ for the conformally invariant scalar equation with $l=2$ and the initial data $\psi(0,x)=(1-x)$, $\psi_\tau(0,x)=0$. The background spacetime parameters are $\sigma=0.7$ (upper left), $\sigma=0.8$ (upper right), $\sigma=0.9$ (lower left), and $\sigma=1$ (lower right), while $\lambda=1$ in all cases.}
    \label{fig:rnads}
\end{figure}

Since the structure of the obtained equation is the same as of the toy-model, we can use the same numerical schemes to evolve it in time. For Eq.\ (\ref{eqn:conformal_RNAdS}) one can define an energy\footnote{We expect it to be bounded from below but since it is used just as a check on numerics, that is not essential.}
\begin{equation*}
    E(t)=\frac{1}{2} \int_0^1 \left[-a_{\tau\tau}\, (\partial_\tau \psi)^2+ a_{xx}\, (\partial_x \psi)^2- a_{0}\,\psi^2 \right]\, dx.
\end{equation*}
Thanks to the lack of $\partial_\tau \psi$ term and due to $a_{\tau x}$ being a constant, $E$ is monotone decreasing as for the toy-model
\begin{equation*}
    E'(t)=\left.\frac{1}{2}a_{\tau x}(\partial_\tau \psi)^2 \right|_{x=0}\leqslant 0.
\end{equation*}
Results of the numerical simulations for various parameters $\sigma$ are presented in Fig.\ \ref{fig:rnads}. Generically the evolution can be divided into three parts: initial behaviour, quasinormal oscillations (which get more distinctive with larger angular numbers), and a monotone decrease. However, the last stage exhibits a power-law decay only in the extremal case, as for the toy-model. For regular black holes the decay is exponential or even absent. To better understand these differences, we calculate quasinormal frequencies with the Leaver method.

Again, the structure of Eq.\ (\ref{eqn:conformal_RNAdS}) lets us use the methods developed in the previous chapter also in this case. However, for this approach to be applicable, one needs to carefully choose the value of $a$ in Eq.\ (\ref{eqn:xa}). In a generic case $f(\rho)$, when expressed via $x$, has four roots: one at $x=0$, one real negative root, and two conjugated complex roots. For our method to converge one needs to choose $a$ in such a way that the three latter zeroes lie outside the circle $|x-1|=1$ in the complex plane. In general the convergence is faster the further the zeroes are from this circle.

\begin{figure}
    \centering
    \includegraphics[width=0.45\textwidth]{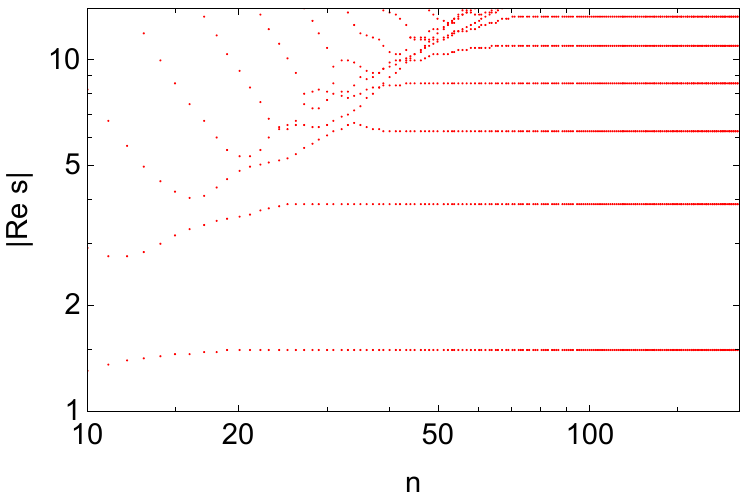}
    \includegraphics[width=0.45\textwidth]{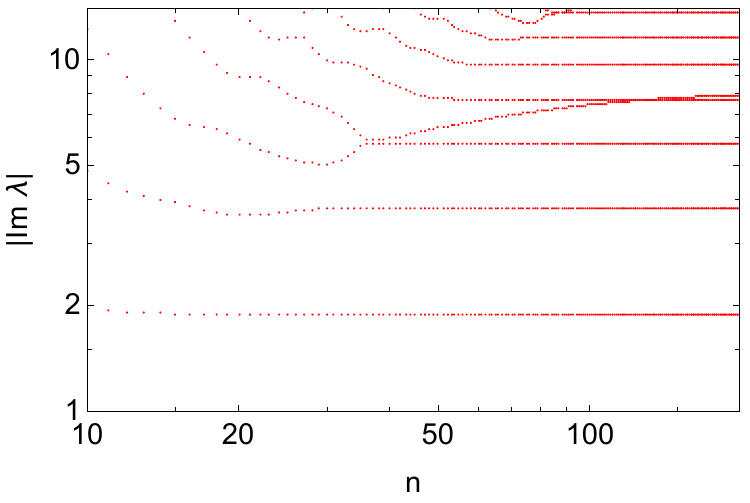}\\
    \includegraphics[width=0.45\textwidth]{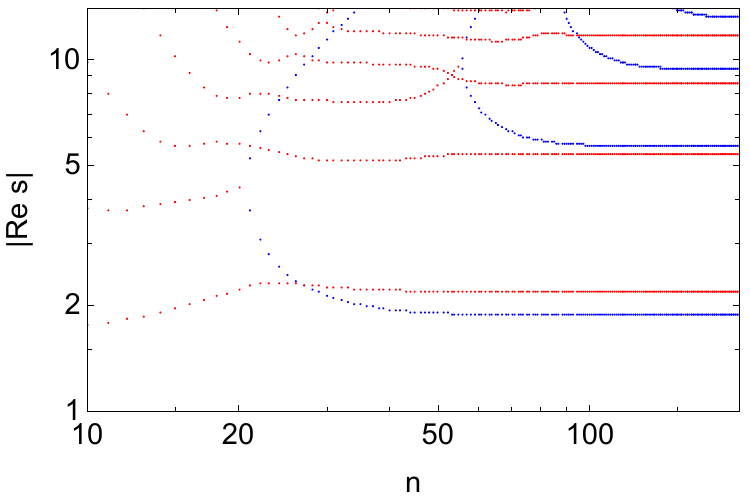}
    \includegraphics[width=0.45\textwidth]{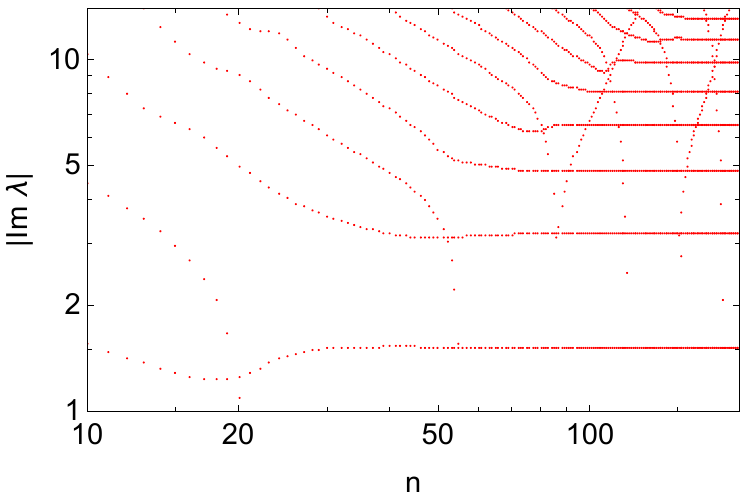}\\
    \includegraphics[width=0.45\textwidth]{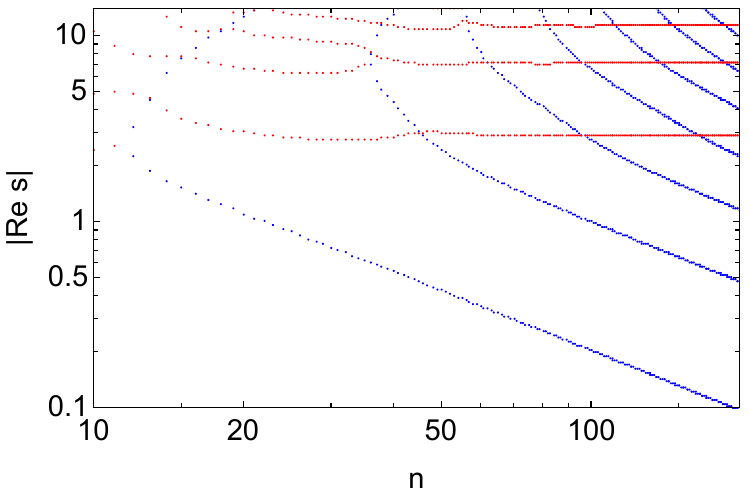}
    \includegraphics[width=0.45\textwidth]{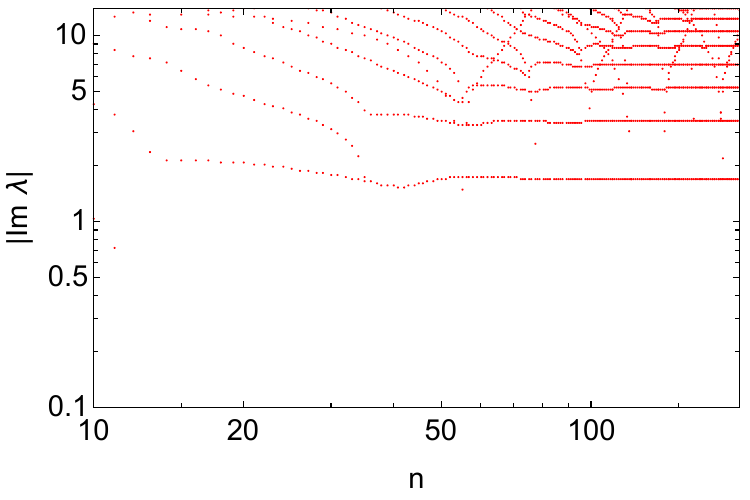}
    \caption{Convergence of real and imaginary parts of oscillatory modes (red) and purely damped modes (blue) for $\lambda=1$, $l=0$, and $\sigma=0$ (left), $\sigma=0.5$ (right), or $\sigma=1$ (bottom) in case of the conformally invariant equation.}
    \label{fig:conv_rnads}
\end{figure}

Figure \ref{fig:conv_rnads} shows how solutions to $H_n=0$ converge for $a=2$, $l=0$, $\lambda=1$, and various $\sigma$. The blue dots denote real solutions (purely damped modes), while the red ones are complex solutions (oscillatory modes). For no charge (Schwarzschild-anti-de Sitter spacetime) only the latter are present. When the charge is non-zero, the purely damped modes appear. As $\sigma$ increases, they get closer to zero but their convergence becomes worse. Finally, in the limit of the extremal black hole we observe similar situation as for the toy model (Fig.\ \ref{fig:conv_toy}): the real solutions become spurious. 

\begin{figure}
    \centering
    \includegraphics[scale=0.7]{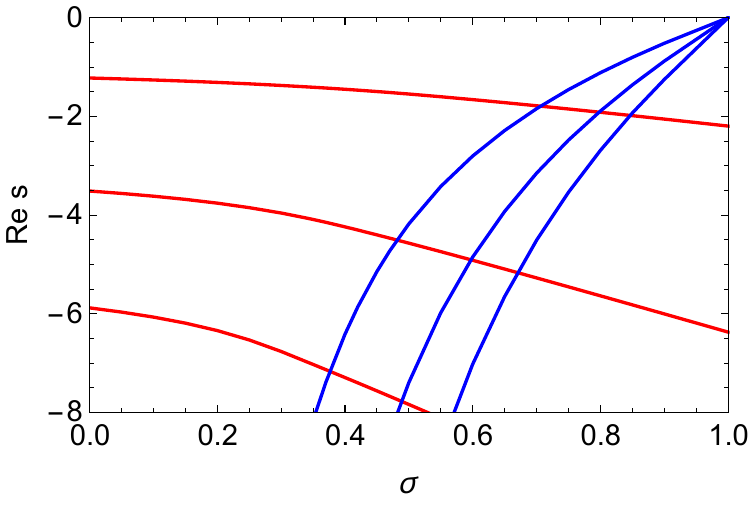}
    \caption{Variation with $\sigma$ of the real parts of three smallest oscillatory modes (blue) and purely damped modes (red) for conformally invariant equation with $\lambda=1$ and $l=2$.}
    \label{fig:exp}
\end{figure}

\begin{figure}
    \centering
    \includegraphics[scale=0.7]{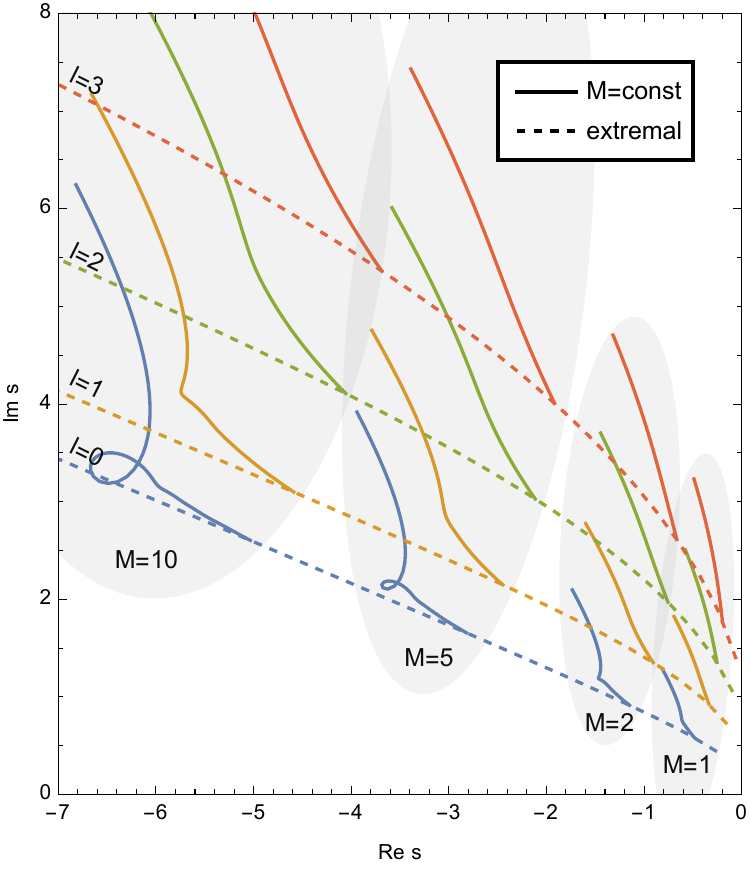}
    \caption{Oscillatory quasinormal frequencies for conformally invariant equation in Reissner-Nordstr\"{o}m-anti-de Sitter spacetime with $\Lambda=-1$ and various masses $M$ and charges $Q$. The dashed lines indicate extremal cases. The solid lines bifurcating from them have constant mass $M$ and are parametrised by decreasing charge.}
    \label{fig:rnads_qnf}
\end{figure}

Figure \ref{fig:exp} shows how real parts of the oscillatory modes and purely damped modes depend on $\sigma$. One can observe that at some point (for $\lambda=1$, $l=2$ it is $\sigma\approx 0.7$) the real part of the lowest decaying mode starts dominating over the lowest oscillatory mode. This transition is reflected in Fig.\ \ref{fig:rnads} by the emergence of the exponential tail. As $\sigma$ grows further, this tail decays. Finally, for $\sigma=1$ all the purely damped modes vanish (they converge to zero) and the tail is described by the power law. This is consistent with the same behaviour that has been proven for the Reissner-Nordstr\"{o}m-de Sitter black hole by Joykutty \cite{Joykutty:2021fgj}. Dependence of the oscillatory modes frequencies on $\sigma$ is less severe and is presented in Fig.\ \ref{fig:rnads_qnf}.


The same approach can be employed also to studies of perturbations of the spacetime. In a typical framework \cite{Cha} they are described by the generalised eigenproblem 
\begin{align*}
    L \psi = V \psi,
\end{align*}
where $V$ is the suitable spherically symmetric potential, depending on the type of the perturbations one studies (for its form in case of the RNAdS spacetime see \cite{Ber03}, let us emphasize here that in RNAdS spacetime with $Q\neq 0$ the electromagnetic and gravitational perturbations are mixed and can be resolved to their axial and polar parts), and
\begin{align*}
    L=\frac{d^2}{dr^2_\ast} +s^2.
\end{align*}
By $r_\ast$ we denote here the tortoise coordinate that for metric (\ref{eqn:RNAdS}) with $r_H=1$ can be defined by
\begin{align*}
    \frac{d\rho}{dr_\ast}=f(\rho).
\end{align*}
This eigenproblem can be obtained from the dynamical equation $\Box \psi + U \psi = 0$, where $\Box$ is a wave operator for the metric (\ref{eqn:RNAdS}) with $r_H=1$ in $(t_H,\rho,\theta,\phi)$ coordinates and
\begin{align*}
    U(\rho)=\left[\frac{f'(\rho)}{\rho}+\frac{l(l+1)}{\rho^2}-\frac{V}{f(\rho)} \right].
\end{align*}
The equation $\Box \psi + U \psi = 0$ can be easily written in coordinates $(\tau, x, \theta, \phi)$. Hence, let us consider a wave equation with a general potential $U$:
\begin{align}\label{eqn:boxV}
    \Box_g \psi + U \psi = 0,
\end{align}
with $g$ denoting metric (\ref{eqn:metric_g}). As we have already pointed out, the wave operator $\Box_g$ contains a non-zero $\partial_\tau$ derivative term. Before we were able to get rid of it using the conformal invariance, however, for general potential $U$ equation (\ref{eqn:boxV}) does not possess this feature. Luckily, in $3+1$ dimensions the wave operator has a useful property that under the conformal transformation $\tilde{g}=\Omega^2 g$, $\tilde{\psi}=\Omega^{-1}\psi$ it behaves like \cite{Wald}
\begin{equation*}
    \Box_{\tilde{g}}\tilde{\psi}=\Omega^{-3} \Box_g \psi -\Omega^{-4} (\Box_g \Omega)\psi.
\end{equation*}
As a result, we can use the same factor $\Omega$ as for the conformally invariant equation to get rid of the $\partial_\tau \psi$ term and the resulting operator will be the same differential operator plus an additional potential term. It leads us to the equivalent wave equation
\begin{align}\label{eqn:pert}
    \Box_{\tilde{g}} \tilde{\psi} +\left[\Omega^{-3} (\Box_g\Omega) + \Omega^{-2} U\right] \tilde{\psi} =0.
\end{align}
This equation is no longer regular since $\Omega^{-3} (\Box_g\Omega)$ behaves like $(1-x)^{-2}$ near $x=1$. However, it does not pose any problem since we are interested in solutions that satisfy Dirichlet condition at this end. Assuming that the solution vanishes at $x=1$ at least linearly together with an additional factor coming from the conformal transformation makes sure that the considered problem is sufficiently regular.

Equation (\ref{eqn:pert}) can be studied for the whole range of charges up to the extremal case with the same methods as discussed before. As an example, in Fig.\ \ref{fig:berti} we show real and imaginary parts of the lowest quasinormal frequencies of the scalar (with $l=0$) and axial (with $l=2$) perturbations (let us point out that due to a different convention real parts of QNFs obtained by us correspond to imaginary parts in their approach, and vice versa). These results regard spacetime with the AdS radius $\ell=1$, and various masses and charges set in such a way that the event horizon is localised at $r_H=5$. The plots are parametrised by the ratio of the charge $Q$ and the extremal charge $Q_{ext}=10\sqrt{19}$ in this setting (in the extremal case $M_{ext}=255$). It lets us compare the results of our approach with the previous results from \cite{Ber03}, where the authors were considering an analogous problem for $Q\leq0.55\, Q_{ext}$, and we find good agreement in this range. In particular, our results agree with relevant numerical values provided in Tables II and III of \cite{Ber03}. In the same work, the authors propose approximating quasinormal frequencies for small charges by a simple polynomial relation. Tables IV and V contain fitted parameters of these polynomials. They strongly depend on the range of the data used to obtain the fit, nevertheless, they agree with our results within reasonable limits.

We expect the same approach to also work for polar perturbations, however, in this case potential $U$ in Eq.\ (\ref{eqn:boxV}) introduces additional poles in the complex plane. Then coordinate $x$ defined in Eq.\ (\ref{eqn:xa}) is not sufficient to move these additional poles outside of the disk of convergence required for the Leaver method, independently of the value of $a$. This feat can be done by considering more complicated compactifications, however, their proper choice seems to heavily depend on the values of $l$, $\lambda$, and $\sigma$. Due to these technical difficulties we decided to focus only on scalar and axial perturbations in this article.

\section{Conclusions} \label{sec:conclusions}
The main goal of this work was to investigate the behaviour of the quasinormal modes of Reissner-Nordstr\"{o}m--AdS black hole as the horizon approaches extremality. We pursued it by using spacelike surfaces intersecting the future horizon. At first we tested this approach on the explicitly solvable toy-model for the waves propagating outside of an extremal black hole. Then we successfully used it to reproduce and extend previous results regarding scalar and axial perturbations of RNAdS black holes \cite{Ber03}. Thanks to the appropriate choice of the slicing we were able to study black holes with any charge, including the extremal case, within a single framework. We observed several interesting phenomena for strongly charged black holes, such as a qualitative change of the least damped mode behaviour for some critical charge value or vanishing of the purely damped modes as black hole becomes extremal.

In Section \ref{sec:rnads} we have pointed out some difficulties arising in the case of polar perturbations in RNAdS spacetime. Overcoming them and comparing the obtained results with previous works \cite{Ber03} would constitute a pretty straightforward extension of our work. Another potential future prospect involves employing similar approach to asymptotically flat black hole spacetimes. Then, by conformal transformation infinity can be identified with an extremal horizon and the methods analogous to the presented above shall be applicable.

\subsection*{Acknowledgements}
We would like to thank Piotr Bizo\'{n}, Maciej Maliborski, and Mengjie Wang for their useful remarks. We are grateful to the Erwin Schr\"odinger International Institute for Mathematics and Physics, where some of this work was undertaken during the Thematic Programme ``Spectral Theory and Mathematical Relativity''. FF acknowledges support by Polish National Science Centre Grant No.\ 2020/36/T/ST2 /00323 and the Austrian Science Fund (FWF) via Project \href{http://doi.org/10.55776/P36455}{P 36455}. 


\end{document}